\documentstyle[twocolumn,times,aps]{revtex}

\begin{document}

\newcommand{\be}{\begin{eqnarray}}
\newcommand{\en}{\end{eqnarray}}
\newcommand{\mc}{\mathcal}
\newcommand{\no}{\nonumber}
\newcommand{\ie}{\mbox{\protect{\it i.\ e.\ }}}
\newcommand{\eg}{\mbox{\protect{\it e.\ g.\ }}}
\newcommand{\cf}{\mbox{\protect{\it c.\ f.\ }}}
\newcommand{\etal}{\mbox\protect{\it et.\ al.\ }}

\twocolumn[
\hsize\textwidth\columnwidth\hsize
\csname@twocolumnfalse\endcsname

\title{
Enhanced Spin Polarization of Conduction Electrons in Ni, 
explained by comparison with Cu
}

\author{K.\ N.\ Altmann and D.\ Y.\ Petrovykh}
\address{
Department of Physics, University of Wisconsin--Madison,
1150 Univ. Av., Madison WI 53706--1390, USA.}
\author{G.\ J.\ Mankey}
\address{
MINT Center, University of Alabama, Box 870209, Tuscaloosa, AL 
35487--0209, USA.}
\author{Nic Shannon}
\address{
Department of Physics, University of Wisconsin--Madison,
1150 Univ. Av., Madison WI 53706--1390, USA.}
\author{N.\ Gilman, M.\ Hochstrasser and R.\ F.\ Willis}
\address{
Department of Physics, Penn State University, University Park, PA 
16802--6300, USA.}
\author{F.\ J.\ Himpsel}
\address{
Department of Physics, University of Wisconsin--Madison,
1150 Univ. Av., Madison WI 53706--1390, USA.}

\date{To be published in Phys. Rev. B, June 2000.}

\maketitle

\begin{abstract}
The spin--split Fermi level crossings of the conduction band in Ni are 
mapped out 
by high--resolution photoemission and compared to the equivalent 
crossing in Cu. The 
area of the quasiparticle peak decreases rapidly below Ef in Ni, but 
not in Cu. 
Majority 
spins have larger spectral weight at Ef than minority spins, thereby 
enhancing the spin--polarization beyond that expected from the 
density of states. A large part of the effect can 
be traced to a rapid variation of the matrix element with {\bf k} at 
the point where 
the s,p--band 
begins to hybridize with the $dz^2$ state. However, it is quite 
possible that the intensity drop 
in Ni is reinforced by a transfer of spectral weight from 
single--particle to many--electron 
excitations. The results suggest that the matrix element should be 
considered for 
explaining the enhanced spin polarization observed for Ni in 
spin--polarized tunneling.
\end{abstract}

]
\narrowtext

\section{Spin--Polarized Currents in Magnetoelectronics}

The rapidly growing field of magnetoelectronics \cite{1,2,3} is 
largely based on the 
manipulation of spin currents that are carried by electrons at the 
Fermi 
level $\epsilon_f$. Examples are the application of giant 
magnetoresistance (GMR) in 
reading heads for hard disks and 
the use of spin--polarized tunneling \cite{4,5} and junction 
magnetoresistance (JMR) for a 
magnetic random access memory (MRAM). Spin--polarized tunneling is 
also being 
explored for high--resolution magnetic imaging by scanning tunneling 
microscopy 
(STM) \cite{6,7,8}. The magnitude of the magnetoresistance increases 
with the spin 
polarization of 
the currents, likewise the magnetic contrast in STM. A variety of 
efforts 
are directed 
towards designing new magnetic materials with higher 
spin--polarization, such 
as half--metallic compounds and nanostructures. For making systematic 
progress one first 
has to 
identify the electronic states that are responsible for the spin 
currents, then 
determine the 
fundamental parameters relevant for spin polarization, and eventually 
apply this 
knowledge to the design of new magnetic materials.

The character of the spin carriers in ferromagnets has been debated 
for some 
time. \cite{9,10,11,12,13,14,15,16,17} 
The initial puzzle has been whether s,p-- or d--electrons dominate 
transport 
properties \cite{9}. 
The s,p--states have high group velocity, but low density of states 
and weak 
magnetism. The d--states carry the magnetic moment and have high 
density of states, 
but their group velocity is low. This dilemma can be resolved by 
looking at a 
realistic band 
structure where a free--electron--like s,p--band hybridizes with the 
magnetic 
d--levels close 
to the Fermi level \cite{9,15}. 
That allows the s,p--band to acquire a significant 
magnetic splitting \cite{14,15,16}. 

The origin of the spin--polarization in is still under intense 
investigation
\cite{10,11,12,13,14,15,16,17}. 
Various mechanisms have been proposed, such as a spin dependence of 
the density of 
states, spin--dependent electron scattering in the bulk and at 
interfaces, and a spin--dependent matrix element. A direct determination of the spin 
polarization from magneto--transport properties is difficult. An 
extensive set of values has been 
reported for spin--polarized tunneling into superconductors \cite{4,5} and Andreev 
reflection at point contacts to 
superconductors \cite{18,19}. The traditional explanation of such 
data has been the imbalance in 
the density of states at $\epsilon_f$ for a magnetically--split 
free--electron 
band \cite{4}.  It has been fairly 
successful for explaining the spin--polarization of Fe, but has failed 
for Ni where the 
observed spin--polarization of 23\%--46\% far exceeds the 6\% 
spin--polarization expected 
from the density of states \cite{4,5,18,19}.  
A variety of more sophisticated approaches have been 
proposed for explaining spin--polarized tunneling 
\cite{10,11,12,13,17}. 
It is highly derirable to achieve 
high spin--polarization in tunneling from Ni alloys, such as permalloy 
(Ni$_{0.8}$Fe$_{0.2}$), since 
permalloy is the most common material in magnetoelectronics.

It is difficult to pinpoint the parameters relevant for achieving 
high spin--polarization from transport data alone. Such measurements integrate 
in {\bf k}--space over the 
Fermi surface and involve additional parameters, such as scattering 
lengths. Angle--resolved photoemission is able to focus onto specific {\bf k}--points 
\cite{20} and to separate out 
scattering lengths \cite{14}. 
Traditionally, this technique has lacked sufficient energy resolution 
for discerning the electronic states that are relevant for transport 
phenomena, that is those 
within a few thermal energies kT of $\epsilon_f$. In our study, the 
energy resolution is 9 meV for 
electrons plus photons, compared to kT=25 meV at room temperature. 
Tunable 
synchrotron radiation allows us to map out the {\bf k} component 
perpendicular to the surface 
independent of the parallel components. As non--magnetic reference 
material we use Cu. It 
has the same crystal structure as the adjacent Ni and a similar band 
topology. The main 
difference is an energy shift of the d--bands, which lie 2 eV lower in 
Cu than in Ni.

Our key observations are two intensity anomalies in the spin--split Ni 
conduction 
band: Below $\epsilon_f$ the band loses intensity very rapidly in Ni 
but not in Cu. Furthermore,  
majority spins have larger photoemission intensity at $\epsilon_f$ 
than minority spins. That creates 
an extra spin--polarization beyond the higher density of states for 
majority spins (which 
enters when integrating {\bf k} over the Fermi surface). Several 
possible explanations are 
explored, such as increased electron scattering below $\epsilon_f$, a 
photoemission matrix element 
that varies rapidly with E and {\bf k}, and a transfer of spectral 
weight from single--electron 
excitations to many--electron excitations. Judging from our comparison 
with Cu and from 
simple matrix element calculations we assign the anomalies in Ni in 
large part to a rapid 
decrease of the matrix element at the point where the s,p--band 
becomes more d--like. Our 
finding suggests that a similar role can be expected from the matrix 
element in other 
phenomena, such as in spin--polarized tunneling.

\section{Manybody States, Spectral Function, and Matrix Element}

For encompassing the possibility of manybody interactions and 
electron scattering 
it is useful to start out with a very general characterization of 
electronic states in solids. 
That can be achieved by a spectral function $A({\bf k},\omega)$ which 
describes the spectral weight as 
a function of energy $E=\hbar \omega$ and momentum ${\bf p}=\hbar {\bf k}$. In a 
band structure model, where only 
single--electron excitations are possible, $A({\bf k},\omega)$ 
consists of sharp d--function peaks. The 
spectral function is far more general than this, however, and can 
describe all the many--electron effects measurable in a photoemission experiment. This 
generality is particularly 
useful for describing correlated electrons in the partially--filled 
3d--shells of ferromagnets. Ni 
exhibits a broad satellite several eV below the single--hole states 
\cite{21,22,23,24,25,26,27,28} which may be viewed 
as a pair of correlated d--holes. The consequences of the two--hole 
satellite for the single--particle excitations in Ni are a reduction of the band width by 40\% 
and a decrease of the 
magnetic splitting by a factor of 2--3 \cite{24,25,26,27}.  
These discrepancies have been associated with 
spectral weight shifting from the single--hole d--bands down to the 
two--hole satellite while 
preserving the center of gravity of the energy spectrum. A 
trading of spectral weight 
between single-- and multi--electron excitations has been observed in 
adsorbates \cite{29} and 
oxides \cite{30}, too. 

Such a manybody effect will be considered as one of two plausibe 
explanations 
for the drop of the photoemission intensity below $\epsilon_f$ in Ni. 
The line of thought is the 
following: The conduction band in Ni is s,p--like above $\epsilon_f$ 
but acquires more d--character 
below $\epsilon_f$ as it starts hybridizing with the d--bands. The 
increasing d--character makes it 
prone to many--electron effects, such as a transfer of spectral weight 
to the two--hole 
satellite. Even the smaller intensity of the minority--spin peak at 
$\epsilon_f$ would find a natural 
explanation in such a scenario, because the minority--spin d--bands lie 
higher in energy and 
hybridize more with the minority--spin conduction band at 
$\epsilon_f$. For assessing this 
hypothesis we use Cu as reference material where many--electron 
effects weak. For 
example, the intensity of the two--hole satellite is 21\% of the 
one--hole states in Ni, but 
only 2.5\% in Cu \cite{23}.

The spectral function $A({\bf k},\omega)$ is related to the 
angle--resolved photoemission 
intensity $I(\omega,{\bf k}$ by a matrix element 
$M({\bf k},\omega)$, that is specific 
to the photoemission process \cite{31} :
\be
I({\bf k},\omega) = A({\bf k},\omega) \times \mid M({\bf k},\omega) 
\mid^2 \times f(\omega)	
\en
The Fermi--Dirac function f(w) gives the occupancy. The spectral 
function itself has the 
form
\be
A({\bf k},\omega) = -\frac{1}{\pi} \Im \left\{
\frac{1}{\omega - \epsilon_0({\bf k}) - \Sigma({\bf k},\omega)}	
\right\}
\en
containing the complex self-energy $\Sigma({\bf k},\omega)$ and the 
electron band dispersion 
$\epsilon_0({\bf k})$. A fundamental sum rule for the spectral 
function implies a trade--off 
between single--electron and many--electron excitations:
\be
\int A(\omega,{\bf k}) d \omega   =   1 
\en
The Fermi--Dirac function is absent, thus requiring an extrapolation 
of photoemission data 
above the Fermi level, or the inclusion of inverse photoemission 
data. One remaining 
piece in Eq. 1 to be determined is the matrix element $M({\bf k})$ 
for single--hole excitations 
from the $\Sigma_1$ band:
\be
M({\bf k}) = \langle Y_{final}({\bf k}) \mid \vec{A}_p \mid 
Y_{initial}({\bf k}) \rangle
\en	     
A is the vector potential of the photon and p the momentum operator. 
We have have 
performed an estimate of M({\bf k}) by using a combined interpolation 
scheme that takes the 
correct band width and splitting of the Ni d--bands into account 
\cite{32}.

\section{The Photoemission Experiment}

In photoemission, the parallel component ${\bf k}_{\parallel}$ is 
conserved and can be 
determined directly from the kinetic energy Ekin and the polar angle 
J of the 
photoelectrons. The 
perpendicular component ${\bf k}_{\perp}$ varies with the photon 
energy hn and can be 
estimated using 
a free electron upper band with an inner potential \cite{20,33} In 
order to obtain a clear--cut 
spectral function we designed the experimental geometry such that it 
isolates a single 
band crossing the Fermi level with a high photoemission cross 
section. This is achieved 
by selecting the $\Sigma_1$ conduction band along the [110] direction 
in {\bf k}--space. 
It crosses $\epsilon_f$ 
about half--way between $\Gamma$ and X and stays as far from the 
d--bands as possible \cite{33}. 
Dipole 
selection rules provide additional selectivity: The choice of 
p--polarized light with the 
electric field vector in the photoemission plane enhances the $\Sigma_1$ band 
due to its even 
mirror symmetry and eliminates d--bands with odd symmetry. As a 
result, the 
photoemission data in Fig. 1 clearly show a single conduction band 
for Cu and a spin--split version of that band for Ni.

As a consistency check we map the same Fermi level crossing from two 
different 
surfaces, the (100) and the (110). For the (100) surface we reach the 
desired location \cite{33} 
with a photon energy $h\nu$=44 eV for Ni ($h\nu$=50 eV for Cu), 
combined with a polar angle 
of about $20 \deg$ along the [011] azimuth. The (110) surface probes the 
same {\bf k}--point with a 
photon energy $h\nu$=27 eV and a polar angle of about 35$\deg$ along 
[$\overline{1}$10 ]. 
For the (100) 
surface one starts at $\Gamma$ for ${\bf k}_{\parallel}$ = 0 and 
reaches X at 
${\bf k}_{\parallel} = \sqrt{2.2}\pi/a = 2.52\AA$ in Ni ($2.46 \AA 
$in Cu). 
For the (110) surface the bands are mapped in reverse, starting at X 
for 
${\bf k}_{\parallel}$=0 and reaching $\Gamma$ at ${\bf k}_{\parallel} 
= \sqrt{2.2}\pi/a$. 
This inverted {\bf k}--scale shows up in Fig. 1 as approximate 
mirror symmetry of the (100) results (left) and the (110) results 
(right).

Comparing the intensities near $\epsilon_f$ one finds opposite 
behavior for Ni and Cu 
(Fig. 1 top versus center). The Ni bands fade very quickly below 
$\epsilon_f$, 
whereas the Cu band 
remains strong and even increases its intensity slightly. Losing 
oscillator strength so 
rapidly in Ni presents a puzzle: Where did the spectral weight go 
that ought to be there 
according to the sum rule in Eq. 3?  Simple technical explanations 
fail. The {\bf k}--acceptance 
of the analyzer would give equal trends for the intensity in Ni and 
Cu, contrary to the 
drop in Ni and increase in Cu. One could argue that the line width 
increases rapidly below 
$\epsilon_f$ in Ni, thereby reducing the peak height. This lifetime 
broadening is due to the rapidly--increasing phase space for creating electron--hole pairs in the 3d 
bands of Ni. This 
hypothesis is discarded by fitting individual energy spectra at 
various {\bf k} with Lorentzians 
in Fig. 2 and plotting the resulting peak areas in Fig. 3. The 
drop--off in Ni remains and 
contrasts with a slight increase in Cu. 

The Lorentzian fit is equivalent to a simplified spectral function
\be
A_0({\bf k}, \omega) =  \frac{1}{\pi} 
\frac{\Gamma({\bf k})}{[w - \epsilon({\bf k})]^2 + \Gamma({\bf k})^2}	
\en
where the self--energy $\Sigma({\bf k},\omega)$ is taken as functions 
of {\bf k} 
only, not of $\omega$ \cite{34}.   
The real part of $\Sigma({\bf k})$ 
is incorporated into the empirical band dispersion 
$\epsilon({\bf k}) = \epsilon_0({\bf k}) + \Re\{\Sigma({\bf k}) \}$. 
The imaginary part $\Gamma({\bf k})= - \Im\{\Sigma({\bf k})\}$ describes a 
Lorentzian lifetime broadening.  A small 
secondary electron background is added for fitting the data, which 
describes 
{\it extrinsic}
energy losses of the photoelectrons on their way out. It consists of 
an integral over the 
Lorentzian line, which is equivalent to a step--like loss function.

In addition to the intensity drop below $\epsilon_f$ there is a 
second anomaly in Ni. The 
area of the minority peak is smaller than that of the majority peak. 
This can be seen best 
from the {\bf k}--distribution of the photoemission intensity at 
$\epsilon_f$ in Fig. 4. 
The area ratio is $I_{\uparrow}/I_{\downarrow} = 1.8$ for Ni(100) and 
$I_{\uparrow}/I_{\downarrow} = 1.2$ for Ni(110). 
According to a single--electron band 
model one would expect very similar spectral weights for the two spin 
components, since 
they are so close together in {\bf k} space. In fact, previous 
photoelectron spectra of the spin--split bands in Ni have usually been fitted 
with equal intensities for the two spins. 
We are able to unambiguously resolve the two components by measuring 
a {\bf k}--distribution 
at $\epsilon_f$, where the lifetime broadening is minimal. This spin 
asymmetry and the 
intensity drop in Ni are not sensitive to adsorbates such as a residual 
gas or a Cu overlayer, establishing them as pure bulk phenomena. 

\section{Possible Explanations for the Anomalies in Ni}

Within the framework established in Eqs. 1--4 there are two places 
where one can 
search for an explanation of the anomalous behavior of Ni relative to 
Cu. These are the 
matrix element $\mid M(w,{\bf k}) \mid^2$ and the spectral function 
$A({\bf k},\omega)$. 
As long as one wants to stay 
within the one--electron picture, the matrix element for excitation of 
single holes is the 
natural starting point. We have applied a combined interpolation 
scheme to the empirical 
band structures of Ni and Cu for obtaining estimates of $\mid 
M(w,{\bf 
k}) \mid^2$ \cite{32}.  The result describes the 
(100) data qualitatively, including the opposite intensity trends for 
Ni and Cu. However, 
quantitative comparisons are fairly sensitive to the exact location 
of $k_F$, and the (110) data 
are not reproduced well. Clearly, more sophisticated calculations of 
the photoemission 
intensity are called for, such as the one--step model with evanescent 
surface wave 
functions. In the absence of quantitative calculations we use 
experimental results for 
explaining how the matrix element modifies the intensities in Ni and 
Cu. The key will be 
a rapid change in the hybridization between the s,p--band and the 
3d--bands with energy.

While the $\Sigma_1$ conduction band corresponds to the s,$p_z$ states 
in Cu, its symmetry 
allows for significant $d_z^2$ character in Ni. The Ni 3$d_z^2$ 
states lie close to $\epsilon_f$ and strongly 
hybridize with the conduction band, whereas the Cu 3d states lie 2 eV 
lower. For finding 
a d--hybridization in Cu comparable to that of Ni one has to look 2 eV 
lower in energy, as 
shown in the bottom panels in Fig. 1. The group velocity, \ie, the 
slope of the Cu 
conduction band is greatly reduced at this point and has become 
comparable to that of the 
Ni. This is the result of an avoided crossing with the $d_z^2$ level 
\cite{33}.  Likewise, the intensity of 
the Cu band decreases strongly at these lower energies, similar to Ni 
below $\epsilon_f$. The same 
situation is surveyed in {\bf k}--space in Fig. 3. Ni and Cu behave 
similar if one shifts the Ni 
data to the point of comparable d--hybridization in Cu, \ie, a shift 
to the left for (100) and 
to the right for (110). A calculation of the matrix element \cite{32} 
for (100) reproduces this 
effect qualitatively. From such similarities between the Ni bands at 
$\epsilon_f$ and the Cu bands 
at 2 eV below $\epsilon_f$ we conclude that the intensity changes in 
Ni are qualitatively consistent 
with a change in the matrix element due to increasing d--hybridization.

The imbalance between the two spin components can be explained in 
similar 
fashion. The minority spin conduction band is more d--like at 
$\epsilon_f$ than its majority partner 
since it hybridizes with the higher--lying minority $d_z^2$--level. 
Therefore, its matrix element 
has decreased more than that of the majority band. The consequence is 
an 
enhanced spin--polarization at $\epsilon_f$ which has implications 
for spin transport phenomena, such as spin--polarized tunneling \cite{4,5} and Andreev reflection at 
ferromagnetic 
point contacts \cite{18,19}.  As 
mentioned above, the traditional density--of--states model fails to 
explain the high spin 
polarization observed in these experiments for Ni. The larger size of 
the majority spin 
Fermi surface in Ni would give only 6\% 
spin polarization, compared to the observed 
23\%--46\%. 
The extra spin--polarization that we find at $\epsilon_f$ enhances the 
density--of--states 
effect and brings theory closer to experiment. For a quantitative 
comparison it will be 
necessary to map this polarization across the whole Fermi surface and 
to replace the 
photoemission matrix element by the tunneling matrix element.

Despite the qualitative success of the single--particle picture one 
ought to consider 
manybody effects in Ni. Excitations of two d--holes are 
well--documented in this 
material \cite{21,22,23,24}.  Can they produce an effect similar to 
the decrease of the matrix element 
with increasing d--hybridization? There is a scenario where two--hole 
excitations steal 
spectral weight from the single--hole band, taking advantage of the 
sum rule in Eq. 3. It is 
not unreasonable to assume that the probability for exciting a pair 
of d--holes increases 
with the d--character of the band. Therefore, the same arguments as in 
the previous two 
paragraphs can be used, where increasing d--character of the band 
gives rise to a 
decreasing matrix element. It appears that only quantitative 
calculations of the matrix 
element can settle this issue. However, there are some interesting 
clues pointing towards a 
contribution of two--hole effects. The intensity drop in Ni is more 
abrupt than that in Cu at 
the point of comparable d--hybridization. This is particularly 
pronounced for the (110) 
surfaces (Fig. 3, right). One might expect a sharper drop--off for a 
two--hole process 
which scales like the square of the d--hybridization. An additional 
clue comes from the 
decreasing strength of two--hole excitations across the Periodic Table 
from Ni to Co and Fe \cite{23}.  
If manybody effects played a role in the spin 
polarization at $\epsilon_f$, their influence 
would gradually fade from Ni to Co and Fe.  
Such a trend would nicely fit the results from 
spin--polarized tunneling, where the (one--electron) 
density--of--states model works best for 
Fe and worst for Ni \cite{4}.

\section{Summary}

In summary, we find a rapid loss of spectral weight in the conduction 
band of Ni 
below the Fermi level $\epsilon_f$, which is opposite to the behavior 
of the analogous band in Cu. 
Possible mechanisms are considered, such as an increasing lifetime 
broadening, the 
single--hole matrix element, and many--hole excitations stealing 
spectral weight from 
single--hole excitations below $\epsilon_f$. The comparison with Cu 
and a simple estimate of the 
matrix element indicate that the single--hole matrix element is able 
to give a qualitative 
explanation. An additional transfer of spectral weight to two--hole 
states is quite possible, 
however. 
The loss of spectral weight is larger for the minority spin band, 
thereby enhancing 
the spin--polarization at $\epsilon_f$. The photoemission data suggest 
that similar enhancements of 
the spin--polarization might occur in magnetotransport and could be 
used in 
magnetoelectronic devices. For example, the spin--polarization 
observed in spin tunneling 
from Ni exceeds the traditional density--of--states model by a factor 
of five. The analogy 
with photoemission suggests that the tunneling matrix element might 
be responsible.

We would like to acknowledge stimulating discussions with J. C. 
Campuzano, R. 
Joynt, A. Chubukov, and J. Allen on many--electron effects and the 
spectral function. This work was supported by the NSF under Award Nos. 
DMR--9815416, DMR--9704196 and DMR--9809423. 
It was conducted at the SRC, which is supported by the NSF under Award 
No. DMR--9531009.

\section*{Figures}

\begin{figure}[tb]
\caption{
E versus ${\bf k}_{\parallel}$ band dispersions of Ni and Cu near the 
Fermi level $\epsilon_f$, obtained by 
parallel detection of E and J. In Ni, the spectral weight drops 
rapidly below $\epsilon_f$ (top), in 
Cu it increases (center).  Cu behaves similar to Ni when looking at 2 
eV lower energies, where the d--hybridization is comparable (bottom). 
The $\Sigma_1$ conduction band is mapped 
from two surfaces at different photon energies in opposite directions 
(left and right). The gray scale represents 
high photoemission intensity as dark.
}
\label{Fig1}
\end{figure}

\begin{figure}[tb]
\caption{
\label{Fig2}
Photoelectron spectra of Ni(110) and Cu(110) versus ${\bf 
k}_{\parallel}$ (symbols, corresponding 
to vertical cuts in Fig. 1). The lines represent a fit by a 
Lorentzian spectral function (Eq. 
5).
}
\end{figure}

\begin{figure}[tb]
\caption{
Spectral weight in Ni and Cu versus ${\bf k}_{\parallel}$, obtained 
from the area of the Lorentzian 
fit in Fig. 2 (Eq. 5). Note the opposite behavior of Ni and Cu near 
$k_F$.
}
\end{figure}

\begin{figure}[tb]
\caption{
Momentum distributions at $\epsilon_f$ for Ni(110) and Ni(100), 
corresponding to horizontal cuts in Fig. 1 (top). The two spin components 
of the $\Sigma_1$ conduction band are 
resolved (arrows). The larger area of the majority spin peak 
indicates an extra spin--polarization, beyond that expected from the 
larger size of the majority spin Fermi surface [13].
}
\label{Fig4}
\end{figure}


\begin{thebibliography}{}

\bibitem{1}  A series of articles on magnetoelectronics is published 
in : Physics Today {\bf 48}, April 1995, p. 24--63.

\bibitem{2}  G.A. Prinz, J. Magn. Magn. Mat. {\bf 200}, 57 (1999).

\bibitem{3}  F.J. Himpsel, J.E. Ortega, G.J. Mankey, and R.F. Willis, 
Advances in Physics {\bf 47}, 511 (1998).

\bibitem{4}  R. Meservey and P.M. Tedrow, Phys. Rep. {\bf 238}, 173 
(1994).

\bibitem{5}  J.S. Moodera, J. Nowak, and R.J.M. van de Veerdonk, 
Phys. 
Rev. Lett. {\bf 80}, 2941 (1998).

\bibitem{6}  M. Bode, M. Getzlaff, and R. Wiesendanger, Phys. Rev. 
Lett. 81, 4256 (1998).

\bibitem{7}  W. Wulfhekel and J. Kirschner, Appl. Phys. Lett. {\bf 
75}, 1944 (1999).

\bibitem{8}  T.--H. Kim, Y.--J. Choi, W.--G. Park, Y. Obukhov, and Y. 
Kuk, Proceedings of STM'99, 
ed. By Y. Kuk, I.W. Lyo, D. Jeon, and S.--I. Park, p. 224 (1999).

\bibitem{9}  M.B. Stearns, J. Magn. Magn. Mat. 5, 167 (1977). 

\bibitem{10}  J.C. Slonczewski, Phys. Rev. B 39, 6995 (1989).

\bibitem{11}  J. Mathon, Phys. Rev. B 56, 11 810 (1997).

\bibitem{12}  J. M. MacLaren, X.--D. Zhang, and W. H. Butler, Phys. 
Rev. B 56, 11 827 (1997).

\bibitem{13}  E.Y. Tsymbal and D.G. Pettifor, J. Phys. Condens. 
Matter 9, L411 (1997).

\bibitem{14}  D. Y. Petrovykh, K. N. Altmann, H. Höchst, M. 
Laubscher, S. Maat, G. J. Mankey, 
and F.J. Himpsel, Appl. Phys. Lett. 73, 3459 (1998). 

\bibitem{15} F. J. Himpsel,  K. N. Altmann, G. J. Mankey, J. E. 
Ortega, and D. Y. Petrovykh, J. 
Magn. Magn. Mat. 200, 456 (1999). 

\bibitem{16}  F. Manghi, V. Bellini, J. Osterwalder, T.J. Kreutz, P. 
Aebi, and C. Arcangeli, Phys. 
Rev. B 59, R10409 (1999).

\bibitem{17} I.I. Mazin, Phys. Rev. Lett. 83, 1427 (1999).

\bibitem{18}  R.J. Soulen, J.M. Byers, M.S. Osofsky, B. Nadgorny, T. 
Ambrose, S.F. Cheng, P.R. 
Broussard, C.T. Tanaka, J. Nowak, J.S. Moodera, A. Barry, and J.M.D. 
Coey, Science 
282, 85 (1998).

\bibitem{19}  S.K. Upadhyay, A. Palanisami, R.N. Louie, and R.A. 
Buhrman, Phys. Rev. Lett. 81, 
3247 (1998).

\bibitem{20}  F.J. Himpsel, Advances in Physics 32, 1 (1983).

\bibitem{21}  C. Guillot, Y. Ballu, J. Paigné., J. Lecante, K. P. 
Jain, P. Thiry, R. Pinchaux, Y. Pétroff, 
and L. M. Falicov, Phys. Rev. Lett. 39, 1632 (1977).

\bibitem{22}  M. Iwan, F.J. Himpsel, and D.E. Eastman, Phys. Rev. 
Lett. 43, 1829 (1979).

\bibitem{23}  F. J. Himpsel, P. Heimann, and D. E. Eastman, J. Appl. 
Phys., 52, 1658 (1981).

\bibitem{24}  F. J. Himpsel, J. A. Knapp, and D. E. Eastman, Phys. 
Rev. B 19, 2919 (1979).

\bibitem{25}  A. Liebsch, Phys. Rev. Lett. 43, 1431 (1979) and Phys. 
Rev. B 23, 5203 (1981).

\bibitem{26}  W. Nolting, W. Borgiel, V. Dose, and Th. Fauster, Phys. 
Rev. B 40, 5015 (1989).

\bibitem{27}  L. C. Davis, J. Appl. Phys. 59, R25 (1986).

\bibitem{28}  M. Springer, F. Aryasetiawan, and K. Karlsson, Phys. 
Rev. Lett. 80, 2389 (1998).

\bibitem{29}  H.--J. Freund, W. Eberhardt, D. Heskett, and E.W. 
Plummer, Phys. Rev. Lett. 50, 768 
(1983).

\bibitem{30}  S. W. Robey, V. E. Henrich, C. Eylem, and B. W. 
Eichhorn, Phys. Rev. B 52, 2395 
(1995).

\bibitem{31}  The sudden approximation is made, which is reasonable 
at high photon energies, such as those in this work. The sum over final 
states in the golden rule expression for photoemission \cite{20} is 
neglected over the narrow E and k range considered here.

\bibitem{32}  G.J. Mankey, unpublished.

\bibitem{33}  For the band topology of Ni along the S axis compare 
Ref. 15, Fig. 1. The location of 
the transitions in k space is given in Fig. 4 of Ref. 15. In this 
{\it 
transverse} geometry 
the ${\bf k}_{\perp}$ broadening induced by the finite escape depth 
is eliminated to first order 
because the broadenend ${\bf k}_{\perp}$ component lies tangent to 
the Fermi surface. 
Synchrotron 
radiation was used with p--polarized light incident 600 from the 
emission direction for 
the (100) surfaces and 500 for the (110) surfaces, with the sample 
normal between the 
photons and electrons. The spectra in Fig. 1 were acquired 
simultaneously over E,{\bf k} 
using a Scienta electron spectrometer. The sample temperature was 200 
K, which 
sharpened the spectral features near $\epsilon_f$ compared to room 
temperature.

\bibitem{34}  Typical models for $\Gamma({\bf k},\omega)$ contain terms 
proportional to $\omega^2$ and $T^2$, in addition to a ${\bf k}^2$ 
term. These are incorporated into an effective $\Gamma({\bf k})$, which 
also takes an experimental broadening into account.

\end{thebibliography}
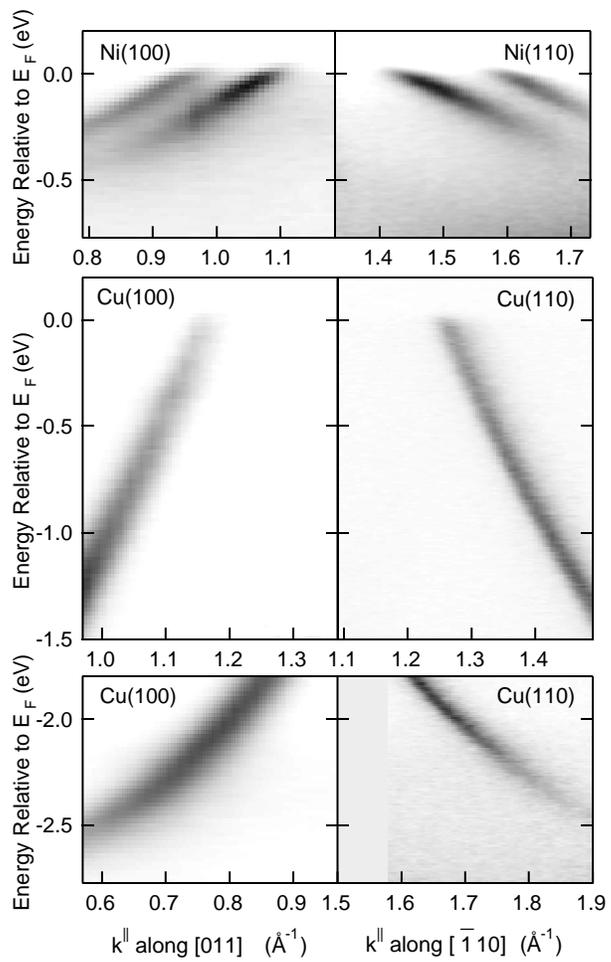
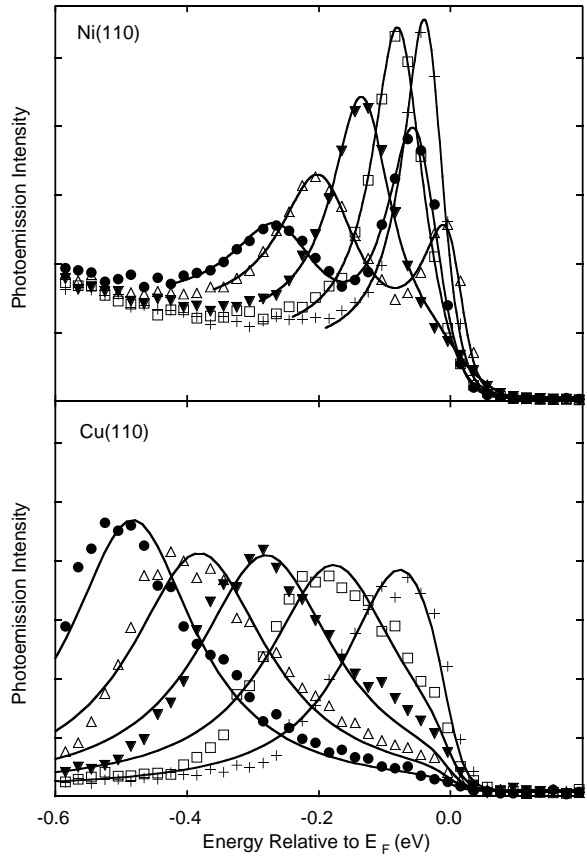
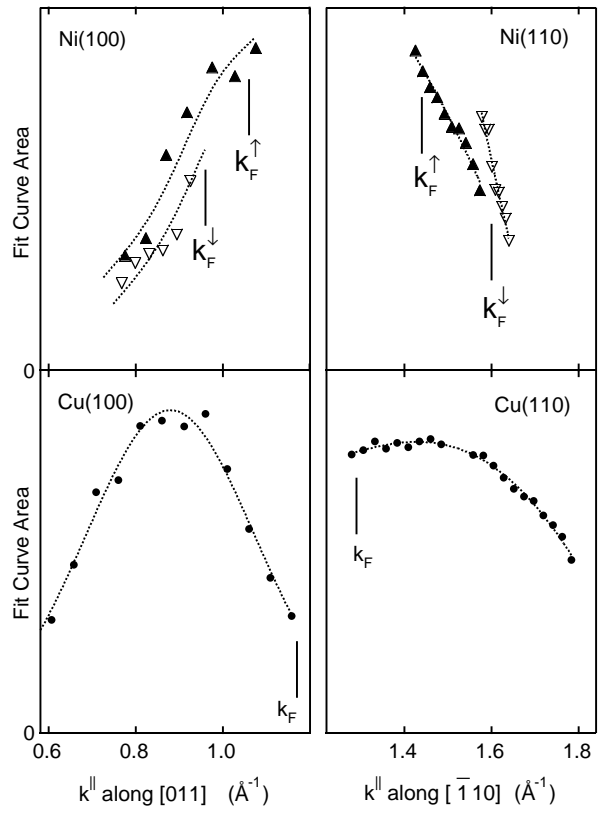
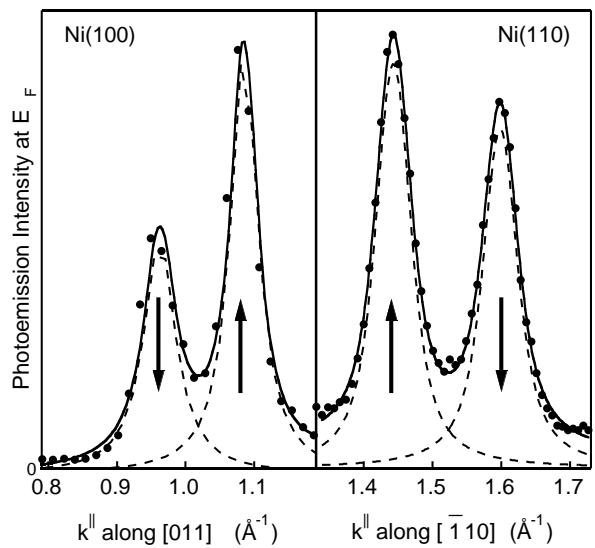
\end{document}